\documentstyle[a4,epsf]{article}

\def\ie{{\rm i.e.\/}\ }

\def\etc{{\rm etc\/}\ }

\def\ZZ{\mbox{\rm Z}\hskip-4pt \mbox{\rm Z}}

\def\CC{\mbox{\rm C}\hskip-6pt \mbox{l} \;}

\def\frac#1#2{{#1\over #2}}

\begin{document}

\begin{titlepage}

\begin{center}

\renewcommand{\thefootnote}{\fnsymbol{footnote}}

\centerline{{\Large On the finite dimensional quantum group}}
\smallskip
\centerline{{\Large $M_3 \oplus (M_{2\vert 1}(\Lambda^2))_0$}}
\vspace{1,5 cm}
\centerline{R. Coquereaux}

\date{September 13, 1996}

\vspace{1.cm}

{\it Instituto Balseiro - Centro Atomico de Bariloche}

{\it CC 439 - 8400 -San Carlos de Bariloche - Rio Negro - Argentina}

\vspace{.5cm}

{\it Centre de Physique Th\'eorique - CNRS - Luminy, Case
907}

{\it F-13288 Marseille Cedex 9 - France }

\end{center}

\begin{abstract}

We describe a few properties of the non semi-simple associative algebra ${\cal H} \doteq  M_3 \oplus
(M_{2\vert 1}(\Lambda^2))_0$, where $\Lambda^2$ is the Grassmann algebra with two
generators. We show that ${\cal H}$ is not only a finite
 dimensional algebra but also a (non co-commutative) Hopf algebra, hence a finite
dimensional quantum group. By selecting a system of explicit 
generators, we show how it is related with the quantum
enveloping algebra of $SL_q(2)$ when the parameter $q$ is a cubic root of 
unity. We describe its indecomposable projective representations as well 
as the irreducible ones. We also comment about the relation between this object and
the theory of modular representations of the group
$SL(2,F_3)$, i.e. the binary tetrahedral group. Finally, we briefly discuss
its relation with the Lorentz group and, as already suggested by 
A.Connes, make a few comments about the possible use of this algebra in a 
modification of the Standard Model of particle physics (the unitary group 
of the semi-simple algebra associated with 
${\cal H}$ is $U(3)\times U(2) \times U(1)$). \end{abstract}

\vspace{2 cm}

\noindent anonymous ftp or gopher: cpt.univ-mrs.fr

\vspace{.5 cm}


\bigskip

CPT-96/P.3388

\vfill {}
\end{titlepage}
%%%%%%%%%%%%%%%%
\newpage
%%%%%%%%%%%%%%%

\centerline{{\Large On the finite dimensional quantum group $M_3 \oplus
(M_{2\vert 1}(\Lambda^2))_0$}}

\section {Introduction}

Quantum groups -- either specialized at roots of unity or not -- have been used many times,
during the last decade, in the physics of integrable models and in conformal theories
\cite{Pasquier-Saleur}. The wish of using such mathematical structures, both nice and new, in
four-dimensional particle physics has triggered the imagination of many 
people in the last years.

When $q$ is a root of unity, there are other interesting objects besides the 
quantized enveloping algebra itself : some of
its  finite dimensional subalgebras or some of its finite
dimensional quotients may still carry a Hopf algebra structure.

We do not need to mention the importance of finite dimensional classical symmetries
in  Physics, but it is our belief that finite dimensional quantum symmetries will turn
out to be also of prime importance in theories of fundamental 
interactions --in this respect, one can already mention the relations
between quantum symmetries of graphs \cite{Ocneanu} and the classification of conformal
field theories \cite{De Francesco-Zuber}.

Such finite dimensional quantum groups are also interesting from the mathematical
point of view because they provide examples of finite dimensional Hopf
algebras which are neither commutative nor co-commutative : they are a kind of
direct ``quantum'' generalization of discrete groups (or, better, 
generalizations of the 
corresponding group algebras). These objects are also
interesting because of their ---still not totally understood--- relations with the
theory of modular representations of algebraic groups \cite{Lusztig1}\cite{Lusztig2}. 

The fact that the semi-simple part of a finite dimensional quotient of 
the quantum algebra $U_q(SL(2,\CC))$,
when $q$ is a primitive cubic root of unity, has a unitary group equal to $U(3)\times
U(2)\times U(1)$ suggests that this finite quantum group could have something to
do with the Standard Model of particle physics. This remark was explicitly made in the framework of non-commutative geometry
by Connes in \cite{Connes1} and more recently in \cite{Connes-Chamseddine}. 
We do not intend, in 
the present paper, to show how to
analyze this finite quantum group along the lines of non-commutative 
geometry (for a very detailed account of 
the Standard Model ``\`a la Connes'', 
not involving quantum groups at all, we refer to the recent papers  \cite{GBV} or \cite{Connes-Chamseddine}).

In order to make use of a symmetry in physics, it is good to be already acquainted
with it. 
It could be tempting to assume that the reader knows already everything about
representation theory of non semi-simple algebras, Jacobson radical, quivers, Hopf
algebras and other niceties belonging to the toolbox of the perfect algebraist but this
would amount to assume that nobody can appreciate the beauty of a tetrahedron before being
acquainted with the properties of the exceptional Lie group $E_6$. Our point of view
is that, since the properties of the algebra ${\cal H}$ can be understood without
using anything more sophisticated than basic multiplication or tensor
products of matrices as well as elementary calculus involving anti-commuting numbers (Grassmann
numbers), it is very useful to study them in this way, at least in a first approach.

We therefore want to present ---in very simple terms--- the rather nice 
finite dimensional algebra of quantum symmetries mentioned before, without assuming from 
the reader any {\it a priori\/} knowledge on quantum groups, general associative 
algebras and the like. We shall therefore define explicitly this finite
dimensional quantum group, as the algebra ${\cal H}\doteq M_3 \oplus (M_{2\vert
1}(\Lambda^2))_0$, where $M_3$ is the set of $3\times 3$ matrices over the complex
numbers, and where $ (M_{2\vert 1}(\Lambda^2))_0$ is the Grassmann envelope of the
associative $\ZZ_2$ graded algebra $M_{2\vert 1}(\CC^2)$, \ie, the even part of its graded tensor product with a Grassmann algebra $\Lambda^2$ with two 
generators. 

The motivation and underlying belief is, of course, that there 
should be some quantum symmetry, hitherto unnoticed, in the Standard 
Model, or, maybe,  in a modification of it, symmetry that would, 
ultimately, cast some light on the puzzle of fermionic families and mass 
matrices. 

The present paper is not only a pedagogical exercise:
although several properties that we shall
describe have been already discussed in the literature (see in particular
\cite{Alekseev},\cite{Suter}), usually using a less elementary language,
 others do not seem to be published. 
 Sections 2 to 5 are supposed to be elementary and self-contained; the last two sections contain a set of less elementary results and unrelated comments.
 
 Finite dimensional quantum groups associated with 
 quantum universal enveloping algebras can be defined for any type of Lie group, when 
 the parameter $q$ is a primitive root of unity. 
 It is possible to give an explicit realization --- in terms of 
 matrices with complex and grassmanian entries --- for the other finite quantum 
 groups of $SL(2)$ type when $q$ is a root of the unity 
 (see \cite{Coquereaux-Ogievetsky}, \cite{Ogievetsky}).
 Following hopes or claims that 
 such algebras can provide interesting physical models, it seems that
 there is some need for a paper explaining 
 the basic properties of the simplest non trival quantum group of this 
 type, namely, when $q^3 = 1$. This is the purpose of the present paper. 
 Construction of a ``generalized'' gauge theory on ${\cal H}$, along the lines of non 
 commutative geometry, is clearly possible but lies beyond the scope of 
 this article.
 
\section {The algebra  ${\cal H}\doteq M_3 \oplus (M_{2\vert 1}(\Lambda^2))_0$}

Let $\Lambda^2$ be the Grassmann algebra over $\CC$ with two generators, 
i.e. the linear span of $\{1,\theta_1,\theta_2,\theta_1 \theta_2 \}$ with arbitrary complex
coefficients, where the  generators satisfy the relations $\theta_1^2 = \theta_2^2 =
0$ and $\theta_1 \theta_2 = - \theta_2 \theta_1$. This algebra has an even part,
generated by $1$ and $\theta_1 \theta_2$ and an odd part generated by $\theta_1$ and
$\theta_2$.
 We call $M_3$ the algebra of $3 \times 3$ matrices over the complex numbers and
$M_{2\vert 1}$ another copy of this algebra that we grade as follows: A matrix $V \in
M_{2\vert 1}$ is called even if it is of the type $$ V = \pmatrix{V_{11} & V_{12} &0
\cr V_{21}& V_{22} &0 \cr 0& 0 &V_{33}} $$ and odd if it is of the type  $$ V =
\pmatrix{0& 0 &V_{13} \cr  0 &0 &V_{23} \cr V_{31} &V_{32} &0 } $$ We call $(M_{2\vert
1}(\Lambda^2))_0$ the Grassmann envelope of $M_{2\vert 1}$ which is 
defined as the even part of the
tensor product of $M_{2\vert 1}$ and $\Lambda^2$, i.e. the space of matrices $3
\times 3$ matrices $V$ with entries $V_{11}$, $V_{12}$, $V_{21}$, $V_{22}$, $V_{33}$,
that are even Grassmann elements (of the kind $\alpha + \beta
\theta_1 \theta_2$)  and entries $V_{13}$, $V_{23}$,$V_{31}$, $V_{32}$
that are odd Grassmann elements (i.e. of the kind $\gamma \theta_1 +
\delta \theta_2$).
We define ${\cal H}$ as
$$ {\cal H} \doteq M_3 \oplus (M_{2\vert 1}(\Lambda^2))_0$$
Explicitly,
$$ {\cal H} =
\pmatrix{
* & * & * \cr
* & * & * \cr
* & * & * } \oplus
\pmatrix{
\alpha_{11} + \beta_{11} \theta_1 \theta_2 &
\alpha_{12} + \beta_{12} \theta_1 \theta_2 &
\gamma_{13} \theta_1 + \delta_{13} \theta_2 \cr 
\alpha_{21} + \beta_{21} \theta_1 \theta_2 &
\alpha_{22} + \beta_{22} \theta_1 \theta_2 &
\gamma_{23} \theta_1 + \delta_{23} \theta_2 \cr
\gamma_{31} \theta_1 +  \delta_{31} \theta_2  &
\gamma_{32} \theta_1 +  \delta_{32} \theta_2  &
\alpha_{33} + \beta_{33} \theta_1 \theta_2 }$$
All entries besides the $\theta$'s are complex numbers (the above
$\oplus$ sign is a direct sum sign: these matrices are $6\times 6$ 
matrices written as a direct sum of two blocks of size $3\times 3$).

It is obvious that this is an associative algebra, with usual matrix
multiplication, of dimension $27$ (just count the number of arbitrary
parameters). ${\cal H}$ is not semi-simple (because of the appearance of
Grassmann numbers in the entries of the matrices) and its semi-simple
part $\overline{{\cal H}}$, given by the direct sum of its block-diagonal
$\theta$-independent parts is equal to the $9+4+1= 14$-dimensional
algebra $\overline{{\cal H}}=M(3,\CC) \oplus M(2,\CC) \oplus \CC$. The 
radical (more precisely the Jacobson radical) $J$ of ${\cal H}$ is
the left-over piece that contains all the Grassmann entries, and only 
the Grassmann entries, so that
$\overline{{\cal H}}={\cal H}/J$. $J$ has therefore dimension $13$.

\section {A system of generators for  ${\cal H}$}

Let $q$ be a primitive cubic root of unity ($q^3=1$). Hence, 
$\overline{q}=q^2=q^{-1}$ and $1 + q + q^2 = 0$. We also set $\lambda = q 
- q^{-1}$. In order to write generators for ${\cal H}$, we need to
consider $6 \times 6$ matrices that have a $(3\times 3) \oplus ((2\vert 
1)\times (2\vert 1))$ block diagonal structure. We introduce elementary 
matrices $E_{ij}$ for the $M(3,\CC)$ part and elementary matrices 
$F_{ij}$ for the $M(2\vert 1,\CC)$ part.

The associative algebra ${\cal H}$ defined previously can be 
generated by the following three matrices
\begin{eqnarray*}
X_+ & = & E_{12} + E_{23} + (1-\theta_1 \theta_2/2) F_{12} + \theta_1 (F_{23} + F_{31}) \\
X_- & = & - E_{21} - E_{32} + (1-\theta_1 \theta_2/2) F_{21} + \theta_2 (F_{13} - F_{32}) \\
K   & = &  q^2 E_{11} + E_{22}+q^{-2} E_{33} + q F_{11} + q^{-1} F_{22} + F_{33}
\end{eqnarray*}
Explicitly, one gets
$$ X_+ \doteq \pmatrix{
\pmatrix{0&1&0\cr 0&0&1\cr 0&0&0} & \pmatrix{} \cr
\pmatrix{} & \pmatrix{
0&1-{\theta_1\theta_2 \over 2}& 0 \cr 
0&0&\theta_1 \cr
\theta_1&0&0}
}$$
$$ X_- \doteq \pmatrix{
\pmatrix{0&0&0\cr -1&0&0\cr 0&-1&0} & \pmatrix{} \cr
\pmatrix{} & \pmatrix{
0&0&\theta_2 \cr 
1-{\theta_1\theta_2 \over 2}&0&0 \cr
0&-\theta_2&0}
}$$
$$
K \doteq  \pmatrix{
\pmatrix{q^2&0&0\cr 0&1&0\cr 0&0&q^{-2}} & \pmatrix{} \cr
\pmatrix{} & \pmatrix{
q&0&0\cr 
0&q^{-1}&0 \cr
0&0&1}
}$$
Performing explicit matrix multiplications or using the relations $E_{ij}E_{jk}=E_{ik}$, $F_{ij}F_{jk}=F_{ik}$ and 
$E_{ij}F_{jk}=F_{ij}E_{jk}=0$,
it is easy to see that the following relations are satisfied:
\begin{eqnarray*}
 X_+^3 = X_-^3 & = & 0, \; K^3 = 1 \\
K X_{\pm} & = & q^{\pm 2} X_{\pm} K \\
\left[X_+,X_-\right] & =  & {K - K^{-1} \over \lambda}
\end{eqnarray*}
It is easy and straightforward to check that the following $27 = 3^3$ 
matrices $\{(X_-^\alpha K^\beta X_+^\gamma)\}_{\alpha,\beta,\gamma \in 
\{0,1,2\}}$ are linearly independent and span ${\cal H}$ as vector space over $\CC$.
This shows that the matrices $X_+, X_-, K$ generate ${\cal H}$ as an algebra.

It is instructive to write these generators in terms of Gell Mann 
matrices, Pauli matrices and $SU(2)$ doublets :

Let $\{\lambda_i\}_{i\in\{1\ldots 8\}}$ denote the Gell Mann matrices (a 
basis for the Lie algebra of $SU(3)$) and $\{\sigma_i\}_{i\in\{1\ldots 3\}}$ 
denote the Pauli matrices (a basis for the Lie algebra of $SU(2)$).
Since we have to use $6\times 6$ matrices, we call $\Lambda_i \doteq {\mbox 
diag}(\lambda_i, 0_{3\times 3})$, $\Sigma_i \doteq {\mbox 
diag}( 0_{3\times 3},\sigma_i,0)$. Therefore
$\Lambda_3 = {\mbox diag}(1,-1,0;0,0,0)$,
$\Lambda_8 = 1/\sqrt 3  \, {\mbox diag}(1,1,-2;0,0,0)$ and we set 
$\Sigma_8 \doteq 1/\sqrt 3 \, {\mbox diag}(0,0,0;1,1,2)$. We shall also
need  the $SU(2)$ doublets\footnote{Warning: our provocative notation refers to
constant matrices, not to fields}
 $(\phi_+, \phi_0) \doteq (F_{13}, F_{23})$ and 
$(\overline{\phi_0}, -\phi_-) \doteq (F_{31}, - F_{32}).$
One can then rewrite the generators $X_\pm$ and $K$ as follows:
\begin{eqnarray*}
X_+ & = & \frac{I_1 + i I_2}{2} (1 - \frac{\theta_1 \theta_2}{2}) + 
(\frac{\lambda_1 + i \lambda_2}{2}) + (\frac{\lambda_6 + i \lambda_7}{2}) +
\theta_1 (\phi_0 + \overline{\phi_0}) \\
X_- & = &\frac{I_1 - i I_2}{2} (1 - \frac{\theta_1 \theta_2}{2}) -
(\frac{\lambda_1 - i \lambda_2}{2}) - (\frac{\lambda_6 - i \lambda_7}{2}) +
\theta_2 (\phi_+ - \phi_-) \\
K   & = &(1/2 + q)\Sigma_3 - (1 + q/2)\Lambda_3  - \frac{\sqrt 3}{2} q 
\Lambda_8 - \frac{\sqrt 3}{2} \Sigma_8
\end{eqnarray*}

Before ending this subsection, we want to note that the matrix 
$C \doteq (q K + q^{-1} K^{-1})/ \lambda^2  + X_- X_+ $ (use 
$\lambda^2=(q-q^{-1})^2=-3$)
commutes with all elements of ${\cal H}$. If we set $q = e^h$,
$K=e^{hH}$ and let $h$ go
to zero (which of course cannot be done when $q$ is a root of unity !), 
the expression of $C$ formally coincides with the usual Casimir operator. 

The  explicit expression of $C$ reads
$$ C = 
\pmatrix{
\pmatrix{-2/3&0&0\cr 0&-2/3&0\cr 0&0&-2/3} & \pmatrix{} \cr
\pmatrix{} & \pmatrix{1/3 - \theta_1 \theta_2&0&0\cr 0&1/3 - \theta_1 
\theta_2&0\cr 0&0&1/3 + \theta_1 \theta_2}}
$$

This operator, when acting by left multiplication on the algebra, 
has two eigenvalues ($-2/3$ and $1/3$) and we see explicitly that the 
eigenspace $C_{-2/3}$ associated with eigenvalue $-2/3$ is isomorphic with the $9$ 
dimensional space $M_3(\CC)$ whereas the eigenspace $C_{1/3}$ associated with eigenvalue $1/3$ 
consists only of nilpotent elements
and coincides with the $13$-dimensional radical $J$ already 
described. In other words, we have $ {\cal H}/C_{1/3} = M_3(\CC) \oplus M_2(\CC) 
\oplus \CC$ and $ {\cal H}/(C_{1/3}\oplus C_{-2/3}) = M_2(\CC) \oplus 
\CC$. We obtain in this way another decomposition of ${\cal H}$ as 
the direct sum of subspaces of dimension $9$, $13$ and $5$ (the 
supplement).

It is useful to consider the following matrix: $p_2 \doteq 1 + 
\lambda^2 C$ because its square projects on the block $M(3,\CC)$ of 
${\cal H}$. The projector is
$(p_2/3)^2  = diag(1,1,1;0,0,0)$.
In the same way, it is useful to consider a matrix that projects on 
the $M_{2\vert 1}$ block of ${\cal H}$. One can use $p_1 \doteq \lambda^2 C 
- 2 = 0_{3\times 3} \oplus (-3) \, diag(1-\theta_1\theta_2,1-\theta_1\theta_2,1+\theta_1\theta_2)$.
 This matrix is not a projector but it nevertheless does the required 
job since it kills the elements of the upper left block. Indeed,

\begin{eqnarray*}
p_1  X_+ & = & 0_{3\times3} \oplus (-3) \{\{ 0, q (1-3(\theta_1 \theta_2)/2, 0\}, \{0, 0,  \theta_1\}, 
 \{ \theta_1, 0, 0\}\} \\ 
p_1  X_- & = & 0_{3\times3} \oplus  (-3) \{\{0, 0, \theta_2\}, \{1 - (3\theta_1  \theta_2)/2, 0, 0\}, 
 \{ 0, -\theta_2, 0\}\} \\ 
p_1   K & = & 0_{3\times3} \oplus (-3) \{
 \{q(1-\theta_1 \theta_2), 0, 0\}, 
 \{ 0,q^{-1}(1- \theta_1 \theta_2), 0\}, 
 \{ 0, 0,1+ \theta_1    \theta_2\}\}
\end{eqnarray*}

 The above properties show that $p_1  p_2^2 = (\lambda^2 C 
- 2)\times (1 + \lambda^2 C)^2 = 0$. 
These two matrices $p_1$ and $p_2$ are very  useful since they allow 
us to express any element of ${\cal H}$ in terms of the generators 
$X_\pm$ and $K$ (something that is for instance needed, if one wants to 
calculate the expression of the coproduct ---see below--- for an arbitrary element of 
${\cal H}$, since the coproduct is usually defined on the generators). 
One can express in this way the $27$ elementary matrices (with or without $\theta$'s)
\begin{eqnarray*}
E_{ij}; & i,j \in &\{1,2,3\} \\ 
A_{ij}\doteq F_{ij}; & (i,j) \in &\{(1,1),(1,2),(2,1),(2,2),(3,3)\}\\ 
B_{ij}=F_{ij} \theta_1 \theta_2 ; & (i,j) \in & \{ (1,1),(1,2),(2,1),(2,2),(3,3)\}\\ 
P_{ij}=F_{ij} \theta_1 ; & (i,j) \in &\{(1,3),(2,3),(3,1),(3,2)\}\\
Q_{ij}=F_{ij} \theta_2 ; & (i,j) \in &\{(1,3),(2,3),(3,1),(3,2)\}
\end{eqnarray*}
 For illustration only, we give :
$E_{11} = {p_2}^2 (X_+)^2 (X_-)^2 /9$ and
$B_{11} = p_1(X_+^2 X_-^2 - X_-^2 X_+^2 + X_- X_+^2 X_-)/6$.

\section{The coproduct on ${\cal H}$}

The fact that  ${\cal H}$, defined in this way, admits a non trivial Hopf algebra 
structure ---in particular a coproduct--- is absolutely not obvious at first sight.

Let us remind the reader that it is the existence of a coproduct
that makes possible to consider tensor products of
representations, exactly as it were a finite group. This is obviously of 
prime importance if one 
has in mind to find some physical interpretation for  ${\cal H}$ and consider 
``many body systems'' (or bound states). For instance, in the case of the 
rotation group, if $J_3$ denotes the third component of angular momentum, 
the coproduct reads $J_3^{total}=\Delta J_3 = J_3 \otimes 1 \oplus 1 \otimes J_3$ and 
this tells us how to calculate the third component of the total angular 
momentum for a system described by the tensor product of two Hilbert 
spaces, namely $J_3^{total} \vert m_1,m_2 \rangle = m_1 \times 1 \vert m_1,m_2 \rangle 
+ 1 \times m_2 \vert m_1,m_2 \rangle = (m_1 + m_2) \vert m_1,m_2 \rangle $. 

We now define directly the coproducts, antipode and counit in ${\cal H}$.
\begin{description}
	\item[Coproduct:]  We define
	    $\Delta X_+ \doteq X_+ \otimes 1 + K \otimes X_+ $,
		$\Delta X_- \doteq X_- \otimes K^{-1} + 1 \otimes X_- $,
        $\Delta  K  \doteq K \otimes K $,
        $\Delta  K^{-1}  \doteq K^{-1} \otimes K^{-1} $.

	\item[Antipode:]  The anti-automorphism $S$ acts as
	    $S 1 = 1$,
	    $S K = K^{-1}$,
	    $S K^{-1} =K$,
     $S X_+ = - K^{-1} X_+$,
	    $S X_- = - X_- K$.
	    As usual, the square of the antipode is an automorphism (and it is, 
	    in this case, a conjugacy by $K^{-1}$, \ie $S^2 u = K^{-1} u K$).

	\item [Co-unit:] The co-unit $\epsilon$ is defined by
		    $\epsilon 1 = 1$,
	    $\epsilon K = 1$,
	    $\epsilon K^{-1} =1$,
     $\epsilon X_+ =0$,
	    $\epsilon X_- = 0$.
\end{description}

Notice that the resulting Hopf algebra is neither commutative nor 
cocommutative.
We now have to check that all expected properties are indeed satisfied.
Here are the main required properties for a Hopf algebra (we do not list the 
usual algebras axioms involving only the multiplication map 
$m : {\cal H} \otimes {\cal H} \mapsto {\cal H}$ and we do not list either the 
axioms involving the antipode).
\begin{description}
	\item[Coproduct:]  $(\Delta \otimes {\mbox id}) \circ \Delta = 
	({\mbox id} \otimes \Delta) \circ \Delta$

	\item[Counit:]  
	$(\epsilon \otimes {\mbox id}) \circ \Delta = {\mbox id}$
	$({\mbox id} \otimes \epsilon) \circ \Delta = {\mbox id}$

	\item[Connecting axiom:]  
	$(m \otimes m) \circ \Sigma_{23} \circ (\Delta \otimes \Delta) =
	\Delta \circ m$ where $\Sigma_{23}$ exchanges the second and third factors in
${\cal H} \otimes {\cal H} \otimes {\cal H} \otimes {\cal H}$.
\end{description}

 The reader may check, in an elementary way, that all these properties 
 are indeed satisfied, either by 
 using the above generators and relations or by using the explicit presentation of 
 ${\cal H}$ given before and explicitly performing the tensor products of matrices.
 For illustration only, we check co-associativity on the generator $X_+$. 
We first compute
\begin{eqnarray*}
((\Delta \otimes {\mbox id}) o \Delta) X_+ & = &
(\Delta \otimes {\mbox id}) (X_+ \otimes 1 + K \otimes X_+) = 
\Delta X_+  \otimes 1+ \Delta K \otimes X_+ \\
  & = & 
  X_+ \otimes 1 \otimes 1 + K \otimes X_+ \otimes 1 + K \otimes K \otimes X_+ 
\end{eqnarray*}
We then compute
\begin{eqnarray*}
	(({\mbox id} \otimes \Delta) o \Delta) X_+ & = & 
	({\mbox id} \otimes \Delta) (X_+ \otimes 1 + K \otimes X_+) =
	X_+ \otimes (1 \otimes 1) + K \otimes (X_+ \otimes 1 + K \otimes X_+) \\
	& = &
	X_+ \otimes 1 \otimes 1 + K \otimes X_+ \otimes 1 + K \otimes K \otimes X_+
\end{eqnarray*}
Both expressions are indeed equal.

To illustrate the non triviality of this result (the existence of a Hopf 
structure on  ${\cal H}$), let us mention, 
for instance, that the algebra $M_2 \oplus \CC$ does not even carry any 
Hopf structure at all (it is known that the only  semi-simple Hopf algebra 
of dimension $5$ is the commutative group algebra defined by 
the cyclic group on five letters). In our case, the presence of the 
$M(3,\CC)$ part is crucial: although  ${\cal H}$ can be written as a direct
sum of two algebras, namely of $M(3,\CC)$ and of $(M_{2\vert 1}(\Lambda^2))_0$, the 
coproduct mixes non trivially the two factors. If one wants to use this 
algebra (or another non co-commutative Hopf algebra) to characterize 
``symmetries'' of some physical system ---for instance in elementary 
particle physics--- one should keep in mind that, in contrast with 
what is done usually in the case of symmetries described by Lie 
algebras, the ``quantum numbers'' will not
usually be additive.

Our explicit description of the algebra allows one to 
compute explicitly the coproduct of an arbitrarily chosen element in 
${\cal H}$.
One has first to express the chosen element in terms of the 
generators $X_\pm$ and $K$ (for that, one may use $p_1$ and $p_2$).
 What is then left to do is a simple calculation 
using the fundamental property of $\Delta$, namely that it is a 
homomorphism of algebras : $\Delta (U  V) = \Delta (U) 
\Delta(V)$ for $U$ and $V$ in ${\cal H}$.
Warning: With the notations given at the end of section $3$, we see that, for
example, $P_{ij}=A_{ij}\theta_1$, however $\Delta P_{ij}$ is {\sl not \/}
equal to $(\Delta A_{ij})\theta_1$, \etc.

In order to appreciate the rather non trivial mixing induced by the 
coproduct, we give --- part of --- the expression $\Delta E_{11}$ (recall that 
$E_{11}$ is an elementary matrix of ``color type'', \ie $3\times 3$, 
containing only a $1$ in position $(1,1)$. We have re-expressed the 
result in terms of elementary matrices $E_{ij}$ ---of color type--- and 
$F_{ij}$ ---of electroweak type.
\begin{eqnarray*}
\Delta E_{11} = & \{- q E_{11}\otimes E_{22} + E_{11}\otimes F_{33} + 
\ldots \}  + & \\
\{
&(1+q) E_{21}\otimes F_{23} - (1+q) E_{31}\otimes F_{13} + \ldots
\} \theta_1 + &\\
\{
&(1+q) E_{12}\otimes F_{32} -  E_{13}\otimes F_{31} + \ldots
\} \theta_2 + &\\
\{
&(1+2q) E_{11} \otimes F_{33} + q E_{22} \otimes F_{22} + \ldots
\}\theta_1 \theta_2 &
\end{eqnarray*}

What is important in this example is not the expression itself (!) but 
the fact that it involves the $E_{ij}$ {\sl and \/} the $F_{ij}$. In a 
sense, one can generate a coupling to the $U(2) \times 
U(1)$ part by building ``bound states'' from the ``color part'' alone.

We can also compute the expression of
the coproduct for a generator of ``electroweak type'',
 like $Q = I_3 + Y/2$ with $I_3 = {\mbox 
diag}(0,0,0;1/2,-1/2,0)$ and $Y={\mbox diag}(0,0,0;-1,-1,-2)$.
A rather long --- but straightforward --- calculation leads (for a two 
body system, \ie  $Q^{tot} \doteq \Delta Q \in {\cal H} \otimes {\cal H})$
to a rather lengthy  result for $Q^{tot}$. The main feature is that
this ``charge'' is not additive : $\Delta Q$ is {\sl not \/} 
equal to $Q \otimes 1 + 1 \otimes Q$ and, moreover, it couples non trivially 
the $M_{2\vert 1}$ part together with the $M_3$ part.

\section {The representation theory of ${\cal H}$}
The theory of complex representations of quantum
groups at root of unity has been worked out by a number of people.
In the case of $SL_q(2,\CC)$, see in particular the articles
by \cite{Pasquier-Saleur}, \cite{Arnaudon} and \cite{Gluschenkov}.
 The study of
representation theory of the finite dimensional algebra ${\cal H}_q$ was
studied by \cite{Suter}. Since our attitude, in the present paper, is 
to study this Hopf algebra without making reference to the general theory 
of quantum groups, we shall not use this last work but describe the representation theory of 
 ${\cal H}$ by using the explicit definition of the algebra given in the 
first section.

Since ${\cal H}$ acts on itself (for instance from the left) one may want to consider the problem of
decomposition of this representation into irreducible or indecomposable 
representations (modules). The problem
is solved by considering separately all the columns defining ${\cal H}
= M_3 \oplus (M_{2\vert 1}(\Lambda^2))_0$ 
as a matrix algebra over a ring. We just
``read'' the following three indecomposable representations from the 
explicit definition of ${\cal H}$ (the
following should be read as ``column vectors''). First of all we have a three dimensional irreducible
representation $M_{st}\doteq (c_1,c_2,c_3)$, (where $c_i$ are complex 
numbers) coming from $M_3$. Notice 
that the three 
columns give equivalent representations. Next we 
have two reducible indecomposable representations (also called ``PIM's'' 
for ``Projective Indecomposable Modules'') coming from the columns of
$(M_{2\vert 1}(\Lambda^2))_0$. Notice that the first two columns give 
equivalent representations (that we call $P_e$), and the last column gives the 
representation $P_o$.
Each of these two PIMS is of dimension $6$. 
$P_o \doteq (\gamma \theta_1 + \delta \theta_2,\gamma' \theta_1 + \delta' \theta_2,
 \alpha + \beta \theta_1 \theta_2) $ and $P_e \doteq (\alpha + \beta \theta_1
\theta_2,
 \alpha' + \beta' \theta_1 \theta_2, \gamma \theta_1 + \delta \theta_2)$.
The notation $M_{st}$ for the three dimensional irreducible representation 
comes from the fact that, in general algebra, such 
modules are called ``Steinberg modules''.  The PIM's are also called ``principal modules''.
Our notation $P_o$ and $P_e$ refers to the fact that, when expressed in 
terms of Grassmann numbers, $P_o$ and $P_e$ are respectively odd and even.

$P_o$ and $P_e$, although indecomposable, are not irreducible : 
submodules (sub-representations)  are immediately found 
by requiring stability of the representation spaces under the left 
multiplication by elements of ${\cal H}$. 

At first sight (see our modifying comment below) one obtains immediately
the following lattice of submodules for the representations $P_o$ and 
$P_e$ (arrows represent inclusions):

$$
{\underline 0}
\rightarrow
{\underline 1}\,
 \vcenter{
\hbox{$\nearrow\raisebox{5pt}{$\underline 3_o$}\searrow$}
\hbox{$\searrow\raisebox{-5pt}{$\underline 3_o'$}\nearrow$}}
\, {\underline 5}
\rightarrow
{\underline 6} = P_o
$$

$$
{\underline 0}
\rightarrow
{\underline 2}\,
 \vcenter{
\hbox{$\nearrow\raisebox{5pt}{$\underline 3_e$}\searrow$}
\hbox{$\searrow\raisebox{-5pt}{$\underline 3_e'$}\nearrow$}}
\, {\underline 4}
\rightarrow
{\underline 6}' = P_e
$$

respectively generated by 
$\underline{5}=(\gamma \theta_1 + \delta \theta_2,\gamma' \theta_1 + \delta' \theta_2,
\beta \theta_1 \theta_2)$, $\underline{3_o} = (\gamma \theta_1,\gamma' \theta_1, \beta \theta_1 \theta_2)$,
$\underline{3_o^\prime} = (\delta \theta_2, + \delta' \theta_2,\beta \theta_1 
\theta_2)$, $\underline{1}=(\beta \theta_1
\theta_2)$ for $P_o$ and by
$\underline{4} = (\beta \theta_1 \theta_2, \beta' \theta_1 \theta_2, \gamma \theta_1 + \delta \theta_2)$,
$\underline{3_e} = (\beta \theta_1 \theta_2, \beta' \theta_1 \theta_2, \gamma \theta_1)$,
$\underline{3_e^\prime} = (\beta \theta_1 \theta_2, \beta' \theta_1 \theta_2, \delta \theta_2)$,
$\underline{2} = (\beta \theta_1 \theta_2, \beta' \theta_1 \theta_2, 0)$ for $P_e$.
Notice that $ W_o \doteq \underline{1} = \underline{3_o} \cap 
\underline{3_o^\prime}$ and that $ W_e \doteq \underline{2} = 
\underline{3_e} \cap 
\underline{3_e^\prime}$. $W_o$ (respectively $W_e$) is called the socle of 
$P_o$ (respectively of $P_e$). The module $\Omega_o \doteq \underline{5}$ 
is the radical of $P_o$ and $\Omega_e \doteq \underline{4}$ is 
the radical of $P_e$. 
 
However, we have forgotten something. Indeed, take $\lambda_1, \lambda_2 \in \CC$, set $\lambda \doteq 
{\lambda_1 \over \lambda_2} \in CP^1$, define $\theta_\lambda = \lambda_1 \theta_1
+ \lambda_2 \theta_2$
and consider the subspace $\underline{3_e^\lambda}$ of $P_e$ 
spanned by $(\beta \theta_1 \theta_2, \beta' \theta_1 
\theta_2, \gamma \theta_\lambda)$ where $\beta, \beta', \gamma$ belong to 
$\CC$.
This subspace is clearly invariant under the left action of ${\cal H}$; 
moreover two representations corresponding to different values of 
$\lambda$ are inequivalent. Appearance of such inequivalent 
representations (for different values of $\lambda$) is related to the fact 
that the group $SL(2,\CC)$ acts by exterior automorphisms on the algebra 
${\cal H}$, since it ``rotates'' the space spanned by $\theta_1$ 
and $\theta_2$. Multiplying $\lambda_1$ and $\lambda_2$ by a 
common scalar multiple amounts to change the coefficient $\gamma$ so that 
this family of representations is indeed parameterized by  $\lambda \doteq 
{\lambda_1 \over \lambda_2} \in CP^1$. The representations $\underline{3_e}$ and $\underline{3_e^\prime}$ described previously 
are just two particular members of this family corresponding to the choices $\lambda = \infty $ and $\lambda = 0$.
A similar phenomenon occurs for submodules of the ``odd'' module $P_o$ 
where we define $3_o^\lambda = (\gamma \theta_\lambda, \gamma' \theta_\lambda 
,\beta \theta_1 \theta_2).$

\begin{figure}[htbp]
\epsfxsize=10cm
$$
    \epsfbox{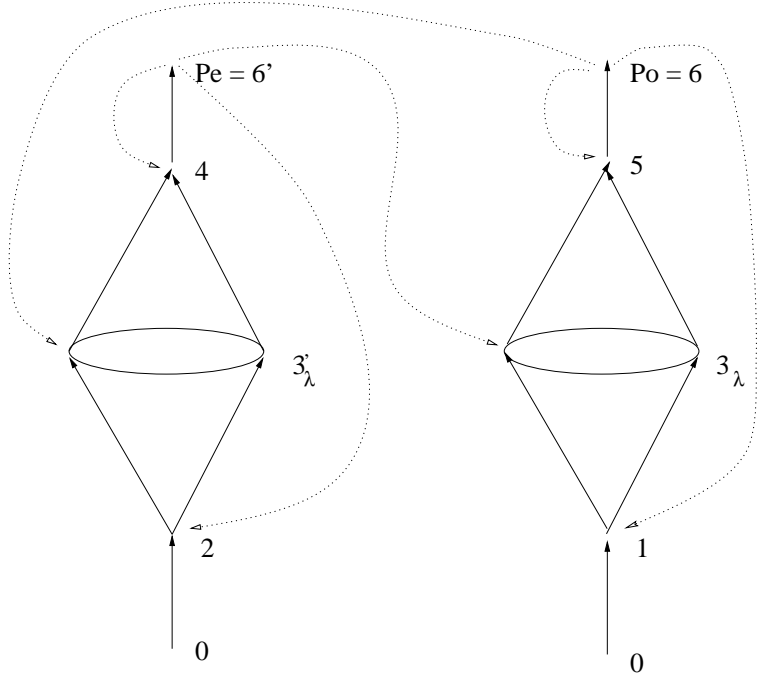}
$$
    \caption{The lattices of submodules for the principal modules of {$\cal H$}}
    \label{fig:lattice}
\end{figure}

The lattices of submodules of $P_o$ and $P_e$ are therefore given by 
figure \ref{fig:lattice}
 
Since we have a totally explicit description of the algebra
and of its lattice of representations, it is easy to continue the 
analysis and to investigate other properties of ${\cal H}$ 
illustrating many other general concepts from the study of non semi-simple associative 
algebras. One can, for instance, study the projective covers of the 
different representations (for completeness sake, this information is 
represented by dashed lines on figure \ref{fig:lattice}), the subfactor 
representations, the quiver of 
the algebra, its Cartan matrix \etc. This, however, would be a bit
technical and more appropriate for a review paper
(see \cite{Coquereaux-Ogievetsky}).

We want only to recall that there exists a one-to-one correspondence 
between irreducible representations of the algebra  ${\cal H}$ and 
the principal modules $M_{st}$, $P_e$ and $P_o$. Irreducible 
representations are obtained from these principal modules by factorizing 
their radical, which amounts to kill the Grassmann ``$\theta$'' variables.
From the above, we see that we obtain in this way three irreducible 
representations : a representation of dimension $3$, $M_{st}$ (it was already irreducible) which 
is a triplet for the unitary group $U(3)$ of the $M_3$ part of ${\cal 
H}$, 
a representation of dimension $2$, $S_e$ (the quotient of $P_e$ by its 
radical) which 
is a doublet for the unitary group $U(2)$ of the $M_{2\vert 1}$ part of ${\cal 
H}$, and finally a representation of dimension $1$, $S_o$ (the quotient of $P_o$ by its 
radical), a $U(1)$ singlet. These are the three irreducible 
representations corresponding to the quotient $\overline{{\cal H}}$ 
of ${\cal H}$ by its Jacobson radical :  (namely $\overline{{\cal H}} 
= \CC \oplus M_2(\CC) \oplus M_3(\CC)$).

The explicit definition given for ${\cal H}$ allows one to compute any tensor 
products of representations and reduce them. We have to consider the
projective 
indecomposable representations ($P_o = \underline{6}$, 
$P_e = \underline{6'}$ and $M_{st}=\underline{{3}}$)
together with the irreducible ones ($S_o = \underline{1}$, $S_e= \underline{2}$
and $M_{st}= \underline{3}$).
Here again appears a mixing between $M_3$ and $M_{2\vert 1})$ 
via the coproduct,  for instance, $\underline{6'}  \times \underline{2} 
\equiv \underline{6}  +  \underline{3} +  \underline{3} $.

\section {Others avatars for ${\cal H}$ and related algebras}

In this paper, we decided to study properties of ${\cal H}$ without using 
any {\it a priori\/} knowledge on quantum groups. Here are nevertheless a 
few  (non elementary) facts, given without proof, that may interest the reader.

Consider the universal enveloping algebra of $SL(2,\CC)$, say $U$. 
Let $U_q$ be the corresponding quantum algebra ($q=e^h$ is a deformation 
parameter) and $X_\pm$, $K$ its generators (not the same as before !).
\begin{itemize}
\item
The first way to construct the finite dimensional Hopf algebra 
${\cal H}$ from the infinite dimensional Hopf algebra $U_q$ is to divide 
it by an ideal (the ideal $I$ defined by the relations 
 given in section $3$)
and check that it is also a Hopf ideal (which means that $\Delta I \subset I\otimes {\cal H} + {\cal H} \otimes I$, $\epsilon(I)=0$ and $S(I) \subset I$).

In the usual $U_q$ ---more precisely in the so called adjoint rational form--- and when $q$ is a cubic root of unity,
the center is generated not only by the Casimir operator but also by the elements ${X_\pm}^3$ and $K^3$.
It is therefore natural to define 
new algebras by dividing the `big' object $U_q$ by an ideal generated by relations of the kind we just considered (remember that $K^3 = 1$ in ${\cal H}$).
Actually, one could as well define new algebras by imposing relations of the
kind $X_\pm^\alpha =0$ and $K^{\pm \gamma}=1$ for integers $\alpha$ and $\gamma$ that are divisible by $3$ (the value of the right hand side, namely $0$ or $1$ is fixed by the existence of a co-unit). The finite dimensional quotient
 ${\cal H}$ therefore appears as a ``minimal'' choice. 
As a matter of fact,
even at the level of ${\cal H}$, defined as before, and without any
 reference to $U_q$, it is convenient to introduce an invertible  square root
 $\hat{K}$ for $K$, hence $K={\hat K}^2$.
 In this way, one obtains a new algebra ${\cal H}^{tot}$ of dimension $2\times 27$ --- just count the number of independent monomials $(X_+^\alpha {\hat K}^\beta X_-^\gamma)$ when $X_\pm^3=0$ and ${\hat K}^6=1$ (this is a PBW basis).
This algebra is quite interesting because its list of representations contains
not only those of ${\cal H}$ but also ``charge conjugate'' representations.
One can also justify, for the quantum enveloping algebra itself,
 and whether $q$ is a root of unity or not, the interest of adding a square root to the generator $K$. Introducing such a square root at the level of $U_q$
 defines the
the so-called simply-connected rational form of the quantum universal enveloping algebra.
The reader should be warned that the algebra ${\cal H}$ or ${\cal H}^{tot}$ is 
sometimes  denoted by $u_q(SL(2,\CC))$ ---with a small $u$--- and
called the ``restricted quantum universal algebra'' (for a reason that will be explained below), but the terminology is not established yet and one should
always look which of $K^3$ or $K^6$ is equal to $1$; for instance, our ${\hat K}$ ---see above--- coincide with the $K$ of \cite{Gluschenkov}.
\item
There exists another construction which is more tricky but maybe more 
profound. Let us start with the following simple observation. Consider the algebra of polynomials with one unknown $x$,  {\sl over the rational numbers\/};
 it can be considered as an algebra
generated by the $x^i$ (to be identified with the
powers of $x$) with relations $x^i x^j = x^{i+j}$. One can make a change of
generators, define the divided powers $x^{[i]}\doteq x^i/i!$ so that
 $x^{[i]} x^{[j]}=\frac {(i+j)!}{i! j!} x^{[i+j]}$ and define the very same algebra
by using the new generators $x^{[i]}$ and the new
relations. The tricky point is that, if we now decide to build an algebra over the finite
field $F_p$ (for instance $\ZZ/3\ZZ$) by taking reduction of coefficients modulo $p$, the two
constructions, with usual powers and divided powers, will lead to two different algebras.
For instance, the relation $x^1 x^2 = x^3$ will be valid in the first algebra (the r.h.s. is
non zero) whereas we obtain $x^{[1]}x^{[2]}=\frac {3!}{1! 2!} x^{[3]} = 0$ if we use divided powers, since $3=0$ in $F_3$. A similar phenomenon appears, in the case of quantum groups when $q$ is a primitive root of unity. 
There are indeed two ways of specializing the $q$ of $U_q$  (see for instance \cite{Chari-Pressley}) to a particular complex value $\epsilon$. One can use the usual ($q$-deformed) generators
and relations, but one can also use $q$-deformed divided powers and corresponding relations
(these relations contain $q$-deformed factorials on the right hand side).
For generic values of $q$, both definitions lead to the same algebra, but when $q$ is a 
primitive root of unity, the algebras are different. More precisely,
we are interested here in the ``restricted integral form'',
 called $U_\epsilon^{res}$ which is obtained as follows.
We assume that $q=\epsilon$, with $\epsilon^3 = 1$,
and start with $U_\epsilon$ considered as an algebra over the cyclotomic field 
$Q(\epsilon)$ (it is obtained by adjoining a cubic root of unity to the 
field $Q$ of rational numbers).
$U_\epsilon^{res}$ is then defined as the subalgebra of $U_\epsilon$ generated  by the elements $(X_\pm)^{(r)}\doteq \frac 
{(X_\pm)^r}{[r]_q!}$, where $[r]_q!$ is the $q$ factorial --- so these 
elements are $q$-deformed divided powers--- and by $K^\pm$. 
Using $[n]_q! \doteq \frac {(q^n-q^{-n})(q^{n-1}-q^{-(n-1)})\ldots(q-q^{-1})}{(q-q^{-1})^n}$,
 we see that $[1]!=1$, $[2]!=-1$ and $[N]_q! = 0$ for $N>2$. This, in turn, implies that,
in the algebra $U_\epsilon^{res}$, $(X_\pm)^3=0$, $K^3$ is central and $K^6 = 1$.
 $U_\epsilon^{res}$
contains a finite dimensional Hopf subalgebra $U_\epsilon^{fin}$, of dimension $2\times 27$ 
over $Q(\epsilon)$, generated by $X_\pm$ and the $K^s$ for 
$s\in{0,1,\ldots,5}$. Since $K^3$ is central, one can also construct a 
quotient (of dimension $27$) by dividing $U_\epsilon^{fin}$ by the 
relation $K^3 - 1$. If we know take arbitrary complex coefficients 
(rather than coefficients belonging to $Q(\epsilon)$), we recover ${\cal 
H}$.
\item
Before ending this section, we want to comment about a rather beautiful 
and mysterious relation with Platonic bodies (actually with the simplest 
of them all, the tetrahedron).
A finite Hopf algebra bearing some strong resemblance with ${\cal H}$ 
was originally defined by \cite{Curtis1},  \cite{Curtis2}  as
the restricted enveloping algebra of a simple Lie algebra over the
finite field $F_p$ with $p$ elements ($p$, a prime). The construction 
goes as follows : 
\begin{enumerate}
	\item  Start with an algebraic group $K$. In our case it 
will be $SL(2,F_3)$. Note that this group, of order $24$ is isomorphic 
with the binary tetrahedral group (the double cover of the $SO(3)$ finite 
subgroup preserving a tetrahedron); the tetrahedron group itself is
 $PSL(2,F_3)$ and is also isomorphic with the alternated group $A_4$. 

	\item  Construct $U_{Z}$, the so-called Chevalley-Kostant 
${\ZZ}$-form of the universal enveloping algebra $U$ for 
the corresponding Lie group $G$, in our case, $G=SL(2,\CC)$. It is 
a subring (over {\ZZ}) of $U$ generated by the divided powers 
$\frac{(X_\pm)^{(r)}}{r!}$.

	\item  Build the hyperalgebra of $SL(2,\CC)$ over $F_p$, namely $U_{F_p} 
	\doteq U_{Z} \otimes_{\ZZ} F_p$ 

	\item  The restricted enveloping algebra $U_{F_p}^{fin}$ is the 
	subalgebra of the hyperalgebra $U_{F_p}$ spanned by the $X_\pm$.
\end{enumerate}

The theory of
restricted enveloping algebras $H_q = U_{F_p}^{fin}(G)$ goes back to 
 \cite{Jacobson2} (see also 
his basic paper on derivations of algebras over a finite field  
\cite{Jacobson1}).
$H_p$ is, in this way, defined for any Lie algebra ${\cal G}$ as a
subring of the corresponding enveloping algebra generated by the
divided powers of the Chevalley generators. The $p$-powers of these 
generators are zero and the obtained algebra is of
dimension $p^{dim {\cal G}}$ over $F_p$. For us, ${\cal G}$ is
$Lie(SL(2,C)$ and $F_p = F_3$ so that $dim \, H_3 = 27$. 
The purpose of defining objects like $H_p$ was historically to study the 
theory of modular representations of finite
Chevalley groups. Although both $H_p$, defined as a restricted enveloping 
algebra over a finite field, and ${\cal H}$, defined as the quotient of
$U_\epsilon^{fin}$ by the relation $(K^3 - 1)$ look like very different 
objects, (the first is an algebra over $F_p$, the
second is over $Q(\epsilon)$ or over $\CC$), it was shown by
\cite{Lusztig1}, \cite{Lusztig2} that there exists a natural bijection between
representation theory over $F_p$ of the first and usual representation theory 
over $\CC$ of the second.
\item
Without entering this deep arithmetical discussion, we want to conclude 
this paragraph by a simple description of the theory of modular 
representations for the binary tetrahedral group $ \Gamma \doteq 
\widetilde{Alt_4} = SL(2,F_3)$. The table of usual (characteristic $0$) characters 
of $\Gamma$ is easy to obtain, for instance from the incidence matrix of 
the extended Dynkin diagram of $E_6^{(1)}$, via the McKay correspondence. 
The dimensions of irreducible representations are simply obtained by 
taking this diagram as a fusion diagram (tensorialization with the 
fundamental of dimension $2$). One obtains in this way the seven inequivalent irreducible representations of $SL(2,F_3)$; see figure \ref{fig:Fusion-E6-ext}.
\begin{figure}[htbp]
\epsfxsize=4cm
$$
\epsfbox{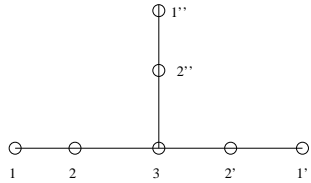}
$$
\caption{Irreducible representations and fusion graph for the binary 
tetrahedral group}
\label{fig:Fusion-E6-ext}
\end{figure}

Modular characters are only interesting in characteristic $2$ and $3$ 
(since primes $2$ and $3$ divide $24$). In 
characteristic $3$, there are only three regular conjugacy classes 
(namely the classes of the identity, minus the identity, and the class of the 
elements of period $4$). Therefore, using Brauer's theory, one can check 
that there are also three irreducible inequivalent modular 
characters, of respective degrees $1$, $2$ and $3$ (like ${\cal H}$ !).
\end{itemize}

\section{Other properties of ${\cal H}$}
\begin{description}

\item[Differential algebras.]

In order to build, in non commutative geometry, a generalized gauge 
theory model, or even something very elementary like the notion
of generalized covariant differential, one needs the following three ingredients.
1)
 An associative algebra ${\cal A}$ (take for instance ${\cal A} = {\cal H}$ or 
${\cal A} = {\cal H} \times C^\infty(R^4))$. 2) A module for ${\cal A}$ 
(choose any one you like).
3) A differential $\ZZ$-graded algebra 
$(\Xi, d)$ that will replace the usual algebra of differential forms (De Rham complex). 
The choice of the last ingredient is not at all unique. For instance, one can 
take for $\Xi$:

1.  The algebra  $\Omega ({\cal A})$ of universal differential form on ${\cal A}$ (one can 
always do so!).
The differential algebra of universal forms on ${\cal H}$ is 
$\Omega({\cal H}) = \bigoplus_{s=0}^{\infty} \Omega^s$
where $\Omega^1 \subset {\cal H} \otimes {\cal H} $ is the kernel 
of the multiplication map, therefore, as a vector space,  
${\cal H} = {{\cal H} \otimes {\cal H} \over \Omega^1}$.
Set $n = dim({\cal H}) = p^3 = 27$, since $p=3$. We see that
$\Omega^1$ is of rank $n-1$ as a ${\cal H}$-module and of 
complex dimension $dim \, \Omega^1 = dim({\cal H})^2 - dim({\cal 
H}) = n(n-1)$. More generally,
$\Omega^s = \Omega^1 \otimes_{\cal H}\Omega^1 \otimes _{\cal H} 
\ldots \otimes_{\cal H}\Omega^1$, so that $\Omega^s$ has rank $(n-1)^s$ as 
a ${\cal H}$ module and is of complex dimension $n(n-1)^s$.

2. The algebra $\Omega_{Der}({\cal A})$ of ${\cal A}$-valued antisymmetric forms on the Lie 
algebra of derivations of ${\cal A}$, which are linear w.r.t. the center 
of ${\cal A}$  ($Der {\cal A} $ is usually not an ${\cal A}$ module). 
This is the choice advocated by \cite{MDV1}.

Understanding the structure of the Lie algebra $Der({\cal H})$ and of its own
representation theory is an interesting subject which we plan to return to in
a separate work. We just recall here a few basic facts. First of all
derivations of $M_n(\CC)$ are all
inner  (they are 
given by commutators). The trace is therefore irrelevant  and  we can identify
 the Lie algebra of derivations of  $M_3(\CC)$  with $Lie(SL(3,\CC))$. By
imposing also a reality condition (hermiticity) one can obtain $Lie(SU(3))$.
Suppose that one defines the algebra $M_2(\CC)\oplus \CC$ in terms
of $3\times 3$ matrices as the linear span of elementary matrices
 $F_{11}, F_{12}, F_{21}, F_{22}$ and $F_{33}$, it is easy to see that
commutators with $F_{13},F_{23},F_{31}$ and $F_{32}$ define derivations that
are not inner since these elements do not belong to  $M_2(\CC)\oplus \CC$, but
they are valued in the module of $3\times 3$ matrices.
In a version of non commutative differential calculus using $\Omega_{Der}$,
such derivations can be related to the notion of Higgs doublets.
 In the case of
 ${\cal H}$, which contains a Grassmann envelope (see first section),
one has also to take into account the fact that the algebra is not
semi-simple since it contains, in particular
 $(M_{2\vert 1}(\CC) \otimes \Lambda^2)_0$. 
Remember that derivations of Grassmann algebras are outer, and that, in
particular, the vector space of {\sl graded\/} derivations of a Grassmann
algebra with two generators can be identified with the Lie superalgebra
$Lie(SL(2\vert 1))$
 whose representation theory (\cite{Marcu}) is known to contain the
representations that are needed to build the Standard Model of electroweak
interactions (although the model is by no means obtained by gauging this
superalgebra! See \cite{Ne'eman-Mieg} and \cite{CES}).

3.  When ${\cal A}$ is the tensor product of $\CC^\infty(M)$ by a 
	 finite dimensional algebra, one can also take $\Xi$ as the 
	 tensor product of the usual De Rham complex, for $M$, times the algebra 
	 of universal forms for the finite geometry. In the case of ${\cal A} =
	 \CC^\infty(M) \otimes (\CC \oplus \CC)$ this was the choice made in 
	 \cite{CEV} (see also \cite{CHS}). Here, keeping in mind applications 
to particle physics, one could take for $\Xi$ the tensor product
 $\Lambda_{DR}\otimes \Omega({\cal H})$, where the first factor refers to the
usual De Rham complex of differential forms over ``space-time''.

4. The algebra $\Omega_{D}({\cal A})$ associated with a $K$-cycle on 
	${\cal A}$, \ie the choice of a Hilbert space and a generalized Dirac operator $D$.
	This is the choice (``spectral triple'')
	 advocated by \cite{Connes2} (and references therein).

	 In the present case, all of the above choices are possible, and also 
	 others, taking into account the existence of twisted derivations, \etc.
	  Since we do not plan here to build any particular physical model, 
	  we stop here our discussion concerning the choice of the 
differential algebra $(\Xi,d)$.

\item[Powers of $SL(2)$ quantum matrices.]

The following observation was made, in $1990$, by \cite{Vokos-Zumino}
and \cite{Corrigan-et-al}: They show that
the $n$-power of a quantum $SU(2)$ matrix with deforming parameter $q$ is
a quantum matrix with deforming parameter $q^n$. This fact was then
recovered and generalized in \cite{Ogievetsky-et-al},
\cite{Ogievetsky-Wess}
. We describe this as follows. Let 
$T \doteq \pmatrix{ a & b \cr c & d}$ be a quantum $SL(2,\CC)$ matrix i.e., with $q$ an arbitrary complex
number, we assume that symbols $a,b,c,d$ obey the six relations $a b = q b a$, $b c =  c b$,
$a c = q c a$, $b d = q d b$, $c d = q d c$ and ${\cal D} \doteq a d - q b c = d a -
q^{-1} b c $, a central element. We define then $a_n , b_n , c_n , d_n$ by 
$$\pmatrix{ a_n & b_n \cr c_n & d_n} \doteq {\pmatrix{ a & b \cr c & d}}^n $$
For instance $a_2 = a^2 + b c$, \etc .
One then shows that the six relations $a_n b_n = q^n b_n a_n$, \etc, are satisfied.
Actually, one proves that $T = q^M$, with $M
 \doteq \pmatrix{ \lambda & \mu \cr \nu  & -\lambda}$,  where $\lambda, \mu$ et
$\nu$ are operators obeying the relations $\left[\mu,\nu \right] = 0$,
 $\left[\lambda,\mu \right] = \mu$ and
 $\left[\lambda,\nu \right] = \nu$. The result concerning powers of quantum matrices follows.

This result implies
immediately that the ``algebra of functions on $SL_{q^r}(2,\CC)$'' is a
subalgebra of the ``algebra of functions on
$SL_{q^s}(2,\CC)$'' as soon as $s$ divides $r$ and that, in particular,
the algebra of functions on the classical group $SL(2,\CC)$ is a
subalgebra of ${ Fun}\,SL_{q}(2,\CC)$ as soon as $q$ is a root of unity. 
This embedding, obtained by using properties of powers of quantum 
matrices is an embedding of algebra but not of coalgebras; this can be seen as
follows (we compare  ${ Fun}\,SL_{q}(2,\CC)$ and  ${ 
Fun}\,SL_{q^2}(2,\CC)$): the coproduct on the algebra spanned by the coordinate functions
$\langle a,b,c,d\rangle$ generating ${\mathsl Fun}(SL_{q}(2,\CC))$ reads, when applied to the
generator $a$,  $\Delta a = a \otimes a + b\otimes c$, \etc. Since $\Delta$ is an algebra
homomorphism   $\Delta a_2 = \Delta a^2 + \Delta(b c) = \Delta a \Delta a + \Delta b \Delta c
$ \ie $\Delta a_2 
= (a \otimes a + b \otimes c)(a \otimes a + b \otimes c) +
 (a  \otimes b + b  \otimes  d)( c  \otimes  a + d  \otimes  c) 
= a^2\otimes a^2 + b a \otimes c a + a b \otimes a c +  b^2 \otimes c^2 + 
a c \otimes b a + a d \otimes b c + b c \otimes d a + b^2 \otimes d c$,
whereas the Hopf algebra ${\mathsl Fun}(SL_{q^2}(2,\CC))$ has {\it another}\/ coproduct,
namely  $\Delta^\prime a_2 = a_2\otimes a_2 + b_2\otimes c_2$ equal to 
$\Delta^\prime a_2 = (a^2 + b c)\otimes (a^2 + b c) + (a b + b d) \otimes (c a + d c)$.
Therefore $\Delta$ and $\Delta^\prime$ are usually different.

Warning: We already mentioned the fact that ${\cal H}_q^{tot}$ (more precisely
 $U_\epsilon^{fin}$) can be considered as a
{\it Hopf}\/ subalgebra of $U_q(SL(2,\CC))$ provided we define it by using divided powers of
the Chevalley
generators. We do not know any relation between this kind of embedding, 
which can be generalized to other $q$-analogues of Lie simple groups \cite{Lusztig1} and
\cite{Lusztig2}) and the algebra embedding mentioned above (which seems 
to be only valid for $SL_q(2)$).

\item[General remarks.]

Embedding of  ${ Fun}\,SL(2,\CC)$  in ${ Fun}\,SL_{q}(2,\CC)$ 
(with $q$ a root of unity) can be visualized as a projection from the
quantum group to the classical one, with a finite quantum group as
``fiber''. This finite quantum group should therefore itself be thought of as
a ``group'' included in $SL_q(2,\CC)$. Despite of the free use of a
terminology borrowed from commutative geometry, note that in the
present situation, spaces have no points (or very few...)!
Morally, one would like to replace the enveloping 
algebra of the Lorentz group $Spin(3,1)=SL(2,\CC)$ by the quantum 
enveloping algebra of $SL_q(2,\CC)$, when $q^3=1$. At the intuitive level 
(and although theses spaces have very few points) one can see the 
classical Lorentz group as a quotient of the quantum $SL_q(2,\CC)$, with 
$q$ a primitive root of unity, by a ``discrete quantum group'' described by 
${\cal H}$. This was the idea advocated in a comment of 
\cite{Connes1}.
We refrain to insist on the obvious similarities between some aspects of representation theory of ${\mathcal H}$ and the Standard Model of elementary particles. We also refrain to insist on the obvious differences$\ldots$
Notice that the finite quantum group
 ${\cal H}$, of dimension $27$ (or ${\cal H}_{tot}$, of dimension $2 \times 27$) is an analogue 
 of the discrete group $\ZZ_2$ that describes the relation 
between the Lorentz group and the spin group (the latter being the 
universal cover of the former): $SO(3,1)=Spin(3,1)/\ZZ_2$, relation which 
is, classically, at the origin of the difference between particles of 
integer and half-integer spin. Here, we have something analogous for 
${\cal H}$. Whether or not one can build a realistic physical model along these lines, 
with non trivial prediction power, remains to be seen. We hope that the 
present contribution may help the interested readers to develop new ideas in 
this direction.

\end{description}

\bigskip
{\Large Acknowledgments} 
\smallskip
 
Many results to be found here came from discussions with Oleg Ogievetsky. 
I want to thank him for many enlightening comments and, in particular, for 
his patience in explaining me the basics concerning 
 representation theory of non
semi-simple associative algebras.

This work was supported, in part, by a grant from the
Instituto Balseiro (Centro Atomico de Bariloche). 
I want to thank everybody there
for their warm hospitality and for providing a peaceful atmosphere which made
possible the writing of these notes.

\eject


\begin{thebibliography}{99}
\bibitem{Ogievetsky} O. Ogievetsky,
Matrix structure of $SL_q(2)$ when $q$ is a root of unity.
CPT-96/P3390
\end{thebibliography}
\end{document}